\documentclass[prc,showpacs,amssymb,floatfix,twocolumn,superscriptaddress]{revtex4}
\usepackage{graphicx}
\usepackage{amsmath,bm,dcolumn}
\usepackage[usenames]{color}
\usepackage{ulem} 
\begin{document}
\title{Shear viscosity of hot nuclear matter by the mean free path method}
\author{D. Q. Fang }\thanks{Email: dqfang@sinap.ac.cn}
\affiliation{Shanghai Institute of Applied Physics, Chinese Academy of Sciences, 
Shanghai 201800, China}
\affiliation{Key Laboratory of Nuclear Radiation and Nuclear Energy Technology, Chinese Academy of Sciences, 
Shanghai 201800, China}
\affiliation{Kavli Institute for Theoretical Physics, Chinese Academy of Sciences, Beijing 100190, China}

\author{Y. G. Ma }
\affiliation{Shanghai Institute of Applied Physics, Chinese Academy of Sciences, 
Shanghai 201800, China}
\affiliation{Key Laboratory of Nuclear Radiation and Nuclear Energy Technology, Chinese Academy of Sciences, 
Shanghai 201800, China}
\affiliation{Kavli Institute for Theoretical Physics, Chinese Academy of Sciences, Beijing 100190, China}
\affiliation{Shanghai Tech University, Shanghai 200031, China}

\author{C. L. Zhou }
\affiliation{Shanghai Institute of Applied Physics, Chinese Academy of Sciences, 
Shanghai 201800, China}
\date{\today}
\begin{abstract}
The shear viscosity of hot nuclear matter is investigated by using the mean 
free path method within the framework of IQMD model. Finite size nuclear sources at 
different density and temperature are initialized based on the Fermi-Dirac distribution. 
The results show that shear viscosity to entropy density ratio decreases 
with the increase of temperature and tends toward a constant value
for $\rho\sim\rho_0$, which is consistent with the previous studies on nuclear matter 
formed during heavy-ion collisions.
At $\rho\sim\frac{1}{2}\rho_0$, a minimum of $\eta/s$ is seen at around $T=10$ MeV
and a maximum of the multiplicity of intermediate mass fragment ($M_{\text{IMF}}$)
is also observed at the same temperature which is an indication 
of the liquid-gas phase transition. 
\end{abstract}

\maketitle

Due to van der Waals nature of nuclear force, liquid-gas phase transition (LGPT) occurs 
in heavy ion collisions (HIC) at energy around hundred MeV/nucleon~\cite{lab1,lab2,Poc,lab3,lab4,Gupta,Bor,Martin}. 
Studies on the phenomena of LGPT and its probes, like fragment size distribution,  
caloric curve, bimodality etc., have become the most important subjects in 
HIC at intermediate energies in the past 
years~\cite{Poc,lab3,Fisher,fluct,Lopez,ma95}.

Viscosity describes fluid's internal resistance 
to flow and may be thought of as a measure of fluid friction. 
In ultra-relativistic HIC, hydrodynamic model 
has been used to study the Quark Gluon Plasma (QGP) phase 
and critical phenomenon~\cite{lab8,lab9,lab10,lab11,lab12,lab13}. 
The investigations show that QGP has very small viscosity over entropy density and behaves 
like a perfect fluid. A few efforts have also been devoted to the study of viscosity 
for nuclear matter formed during HIC at intermediate energies~\cite{lab14,lab15,shi,pal,Shlomo,lisx,zhou1,zhou2,xujun1,xujun2}.
It is found that the shear viscosity to entropy density 
ratio for matter such as H$_2$O, He and Ne2 has a minimum at or near the critical point 
of phase transition~\cite{lab9,lab5}. This is an empirical observation for many kinds of substances.
A lower bound of this ratio ($\eta/s\geq 1/4\pi$) is 
speculated to be valid universally according to certain gauge theory 
(Kovtun-Son-Starinets (KSS) bound)~\cite{lab6,lab7}.
By using the isospin-dependent quantum molecular dynamics (IQMD) model, 
we have studied the shear viscosity of finite size hot nuclear matter ($T>0$) at 
different densities and also the temperature dependence of $\eta/s$. 
The main purpose of this work is to see whether $\eta/s$ exhibits a minimum at the temperature 
of liquid-gas phase transition for nuclear matter.

The quantum molecular dynamics (QMD) model is a many-body theory which describes 
collisions between two nuclei~\cite{lab20,lab21}. Each nucleon $i$ in the nuclei is 
represented by a Gaussian wave packet with definite width ($L=2.16$ fm$^{2}$ in the 
present study) centered around the mean position and the mean momentum.
Then the nucleons propagate in the effective nuclear mean field given by
\begin{eqnarray}
U(\rho,\tau_{z}) =&\alpha(\frac{\rho}{\rho_{0}})+\beta(\frac{\rho}{\rho_{0}})^{\gamma}
                   +\frac{1}{2}(1-\tau_{z})V_{c} \nonumber \\
                  & +C_{\mathrm{sym}}\frac{(\rho_{n}-\rho_{p})}{\rho_{0}}\tau_{z} +U^\mathrm{Yuk}, 
\end{eqnarray}
with the normal nuclear matter density $\rho_{0}=0.16$~fm$^{-3}$.
$\rho$, $\rho_{n}$ and $\rho_{p}$ are the total, neutron and
proton density, respectively. $\tau_{z}$ is the $z$-th component of
the isospin degree of freedom, which equals $1$ for neutron or $-1$
for proton. The coefficients $\alpha$, $\beta$ and
$\gamma$ are parameters for the nuclear equation of state (EOS).
$C_\mathrm{sym}$($=32$~MeV) is the symmetry energy strength. In the present work, we take
$\alpha=-356$ MeV, $\beta$ = 303 MeV and $\gamma = 1.17$ which
correspond to the so-called soft EOS with the incompressibility of $K=200$~MeV~\cite{lab20}. 
$V_\mathrm{c}$ is the
Coulomb potential and $U^\mathrm{Yuk}$ is Yukawa (surface) potential. 
The dynamics of HIC at intermediate
energies is governed mainly by three components: the mean field,
two-body collisions, and Pauli blocking. Therefore, for an
isospin-dependent reaction model, it is important to include
isospin degrees of freedom in the above three components. 
The IQMD model is based on QMD model with consideration 
of the isospin effects in all these processes~\cite{caoxg,taoc}.

\begin{figure}[t]
\centering\includegraphics[width=4.5cm]{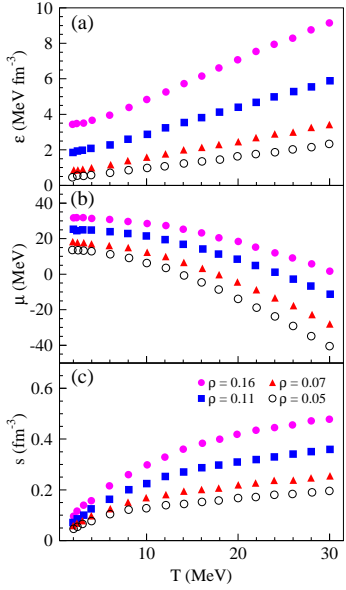}
\caption{(Color online) Temperature dependence of energy density (a), 
chemical potential per nucleon (b)
and entropy density (c) for nuclear sources at different densities (the unit of $\rho$ is fm$^{-3}$).}
\label{fig1}
\end{figure}

Usually the IQMD model is used to simulate the colliding process between two nuclei.
But in the present study, only single nuclear source at $T>0$ is simulated. 
The evolution process and thermal properties of this hot nuclear source are studied.
The nuclear source with $A$ nucleons is initialized using uniform density distribution with
the radius given by $r_0A^{1/3}$ ($r_0$ is the radius parameter).
Then the initial coordinate of nucleons in the source are obtained by 
the Monte Carlo sampling method. 
In the IQMD model, the nucleon radial density can be written as
\begin{eqnarray}
\lefteqn{\rho(r)=\sum_{i}\frac{1}{(2\pi L)^{3/2}}\exp
(-\frac{r^{2}+r_{i}^{2}}{2L})\frac{L}{2rr_{i}} {} }
    \nonumber\\
&& \qquad {}\times \left[\exp(\frac{rr_{i}}{L})
-\exp(-\frac{rr_{i}}{L})\right],
\label{eqrho}
\end{eqnarray}
with the summation over all nucleons. In the usual QMD study, the initial state 
of nuclei is in the ground state with $T=0$. The momentum distribution
of nucleon is generated by means of the local Fermi gas approximation
with the Fermi momentum calculated by $P_F^{i}(\vec{\mathbf{r}}) = \hbar \left[
3\pi^{2}\rho_{i}(\vec{\mathbf{r}}) \right]^{1/3}$,
$\rho_{i}$ is the local density of neutron (i=n) or proton (i=p). 
To study hot nuclear source with $T>0$, the initial momentum distribution
of nucleon is determined by the Fermi-Dirac distribution at finite temperature, 
\begin{equation}
n(e_k)=\frac{g(e_k)}{e^{\frac{e_k-\mu}{T}}+1}, 
\end{equation}
where 
$g(e_k)=\frac{V}{2\pi^2}(\frac{2m}{\hbar^2})^{\frac{3}{2}}\sqrt{e_k}$, 
is the state density for the state  with kinetic energy $e_k=\frac{p^2}{2m}$, 
$p$ is the momentum of nucleon. $V$ is the volume of the source. 
$\mu_i$ is the chemical potential
of nucleon which is determined by the following implicit equation
\begin{equation}
\frac{1}{2\pi^2}(\frac{2m}{\hbar^2})^{\frac{3}{2}}\int_0^{\infty} 
\frac{\sqrt{e_k}}{e^{\frac{e_k-\mu_i}{T}}+1}de_k=\rho_{i}.  
\label{equche}
\end{equation}

\begin{figure}[t]
\centering\includegraphics[width=4.5cm]{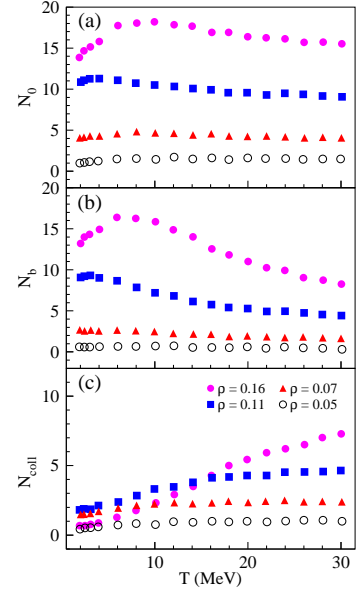}
\caption{(Color online) Temperature dependence of the number of possible collision (a), 
Pauli blocked collision (b) and real collision (c) per fm/c for nuclear sources 
at different densities. For details see the text.}
\label{fig2}
\end{figure}

In this work, the nuclear source with 50 protons  and 62 neutrons ($^{112}$Sn) 
is chosen. The simulated temperature is $0<T<30$ MeV. To study the density 
dependence, different radius parameters ($r_0$) are used to obtain different nuclear densities. 
The thermal properties could be calculated from the phase space information
of the nucleons~\cite{muronga,lisx}. The mean kinetic energy density is calculated by
\begin{equation}
\varepsilon=\frac{1}{A}\sum_{i=1}^{A} e_k^i \rho_i, 
\end{equation}
where $e_k^i$ and  $\rho_i$ are the kinetic energy and local density of the $i$-th nucleon. 
The mean pressure of the system is calculated by 
$P=\frac{2}{3}\varepsilon$. The chemical potential of each nucleon is determined 
by Eq.(\ref{equche}). The entropy density is calculated by
\begin{equation}
s=\frac{\varepsilon+P-\mu\rho}{T}, 
\end{equation}
where $\mu$ is the mean chemical potential of the nucleon, $\rho$ is the mean total nucleon
density.

\begin{figure}[t]
\centering\includegraphics[width=5.cm]{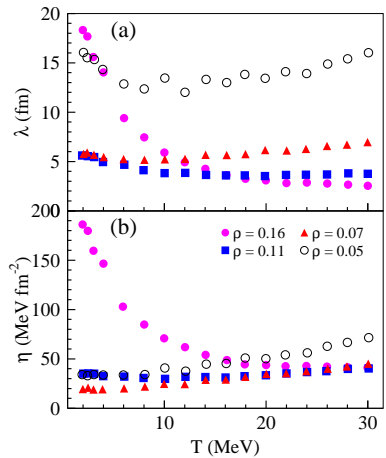}
\caption{(Color online) Temperature dependence of 
the mean free path (a) and the shear viscosity (b) for nuclear sources 
at different densities.}
\label{fig3}
\end{figure}

In the IQMD model, the temperature and density of the system changes with time when
the nucleon propagates in the mean field. The density distribution at any time $t$ could be 
calculated by Eq.(\ref{eqrho}), but the exact temperature $T$ at time $t$ is quite difficult
 to be extracted. 
To investigate the shear viscosity of nuclear matter at definite temperature and density, 
thermal properties and $\eta$ are extracted from the phase space of nucleons at the 
very early time ($1<t<5$ fm/c). During this period of time, the temperature of the system 
is almost the same as the initial value which is given by input. In Fig.~\ref{fig1}(a)-(c), the temperature 
dependence of energy density, chemical potential per nucleon and entropy density 
are shown. As expected, the energy density increases with $T$.
Since the nuclear source studied is almost symmetric, the mean chemical potential of 
neutron and proton is presented in the figure. 
$\mu$ is close to  the Fermi energy when $T$ is close to 0. Due to the Fermi-Dirac
distribution for the momentum of nucleon, $\mu$ decreases
with the increase of $T$ which makes more particles occupy high energy states.
The entropy density increases with $T$, indicating that the system becomes more disorder
with the increase of temperature.
The energy density, chemical potential and entropy density are
proportional to the nuclear density as shown in the three panels. 

In Refs.~\cite{lisx,zhou1}, we have studied $\eta$ of nuclear matter formed 
during HIC processes using the Green-Kubo formula
which employs the linear-response theory to relate the transport coefficients 
as correlations of dissipative fluxes~\cite{kubo} and also a parameterized function
by P. Danielewicz~\cite{lab15,zhou2}. On the other hand, the classical kinetic theory
relates $\eta$ of the system with the mean free path of the particle.
In the kinetic theory~\cite{huang}, 
\begin{equation}
\eta=\frac{1}{3}\rho mv\lambda, 
\end{equation}
where $\rho$, $m$ and $v$ are the density,  
mass and velocity of the nucleon. $\lambda$ is the mean free path
of nucleon which can be determined by  
\begin{equation}
\lambda=\frac{A}{2N_{\text{coll}}}v, 
\end{equation}
with $N_{\text{coll}}$ being the collision number per fm/c (or collision frequency).

The possible collision, Pauli blocked collision and real collision 
numbers per fm/c against temperature are shown in Fig.~\ref{fig2}.
The possible collisions ($N_0$) in the IQMD simulation is determined
directly by the nucleon-nucleon cross section. Among these collisions, 
some collisions will really happen but some collisions will not happen at all.
The collision not happened is the Pauli blocked collision ($N_b$). 
While the real happened collision is denoted by $N_{\text{coll}}$.
From the definition,  we have $N_{\text{coll}}=N_0-N_b$.
At low temperature, the Pauli blocking effect is large due to
the small phase space (momentum space), especially in the high density case.
At high temperature, nucleon can occupy high momentum state which increases
the phase space of the system. In this case, the Pauli blocking effect decreases. 
The effect of Pauli blocking is clearly seen from Fig.~\ref{fig2}(a)-(c), which will greatly 
affect the mean free path of nucleon and also the shear viscosity, especially at low temperature.
The obtained mean free path is given in Fig.~\ref{fig3} (a).
When the density is fixed, the mean free path is proportional to the velocity (temperature)
but inversely proportional to the collision frequency. At high density like 
$\rho=0.16$ fm$^{-3}$, the collision frequency increases very quickly with the temperature. 
The mean free path decreases with $T$ even though the velocity increases with the temperature.
While at low density like $\rho=0.05$ fm$^{-3}$, the collision frequency is almost 
constant in comparison with the high density case, thus the mean free path increases 
with $T$. For the other two densities, the decrease of collision frequency and increase
of velocity is comparable, so the mean free path doesn't change too much with the increase of $T$.

The shear viscosity is shown in Fig.~\ref{fig3} (b). 
At density $\rho=0.16$ fm$^{-3}$, $\eta$ decreases with $T$ and is saturated 
to a value around 50 MeV fm$^{-2}$ that is similar to the 
parameterized formula of $\eta(T)$~\cite{lab15} as well as the Green-Kubo method~\cite{zhou1}. 
At low densities, $\eta$ decrease with $T$ first but increase with 
$T$ when $T$ is larger than $8\sim10$ MeV.

\begin{figure}[t]
\centering\includegraphics[width=5.5cm]{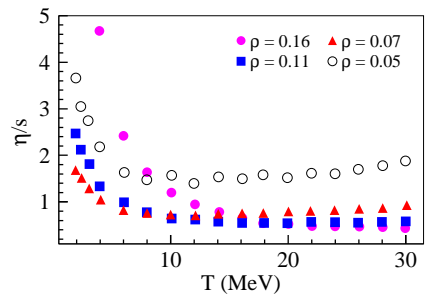}
\caption{(Color online) 
Temperature dependence of $\eta/s$ for nuclear sources 
at different densities.}
\label{fig4}
\end{figure}

$\eta/s$ is shown in Fig.~\ref{fig4}.
At densities $\rho=0.16, 0.11$ fm$^{-3}$, $\eta/s$ decreases with the increase of $T$ and becomes 
saturated to a value round 0.75 (about $7-8$ times of KSS bound) when $T>20$ MeV. 
But at low densities of $\rho=0.07$ and $0.05$ fm$^{-3}$, a minimum is 
seen at around 10 MeV. From the general relation between $\eta/s$ and $T$, this minimum 
might have connection with the liquid-gas phase transition point of nuclear matter. 
In order to check this conclusion, 
multiplicity of intermediate mass fragment (IMF) is investigated since it is a widely studied  
observable for LGPT.

\begin{figure}[t]
\centering\includegraphics[width=5.5cm]{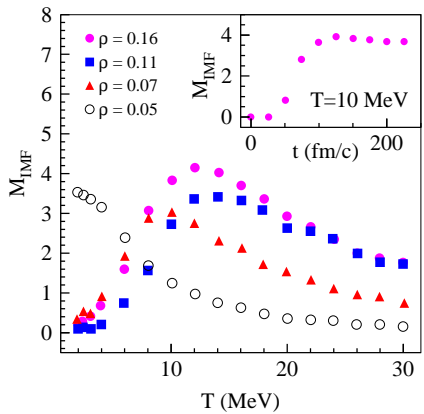}
\caption{(Color online) 
Temperature dependence of the multiplicity of IMF ($M_{\text{IMF}}$) after the 
freeze-out of the system for nuclear sources at different densities.
Inset is the time evolution of $M_{\text{IMF}}$ at $T=10$ MeV for $\rho=0.16$ fm$^{-3}$.
For details see the text.}
\label{fig5}
\end{figure}

To obtain the fragment distribution, evolution of the hot nuclear source is studied.
After the initialization, the nucleons propagate in the mean field with no boundary condition.
It means that the system will expand and freeze-out by emitting nucleon and fragments.
The nuclear fragments are constructed by using the coalescence model, 
in which nucleons with relative momentum smaller than 
$300$~MeV/c and relative distance smaller than
$3.5$~fm will be combined into a cluster.
A fragment with its charge number ($Z$) fulfilling $3\le Z\le Z_{\text{src}}$
is defined as IMF, where $Z_{\text{src}}$ is the charge number of the nuclear source.
From the time evolution of the fragment distribution, it is found that the yields of 
each fragment become stable after 100 fm/c for almost all temperature $T$. 
It indicates that the system is freezed-out after this time as shown by the time 
dependence of the multiplicity of IMF ($M_{\text{IMF}}$) at $T=10$ MeV 
in the inset of Fig.~\ref{fig5}. $M_{\text{IMF}}$ against $T$ at 150 fm/c is shown 
in Fig.~\ref{fig5}. 
At $\rho=0.07$ fm$^{-3}$, a maximum of $M_{\text{IMF}}$ is seen at 10 MeV which is 
almost the same as the temperature of the minimum of $\eta/s$ at this density as 
shown in Fig.~\ref{fig4}. This  result indicates that $\eta/s$ has a minimum 
value at $T\sim10$ MeV when the density is around $\frac{1}{2}\rho_0$. 
At the same temperature, $M_{\text{IMF}}$ of the system after freezing-out 
takes a maximum value which is the signal of liquid-gas phase transition~\cite{Ma95}.
The phase transition temperature is consistent with other studies~\cite{lab4,Song91}.
While for higher densities $\rho=0.16, 0.11$ fm$^{-3}$ ,  no minimum of $\eta/s$ is seen
but instead $\eta/s$ approaches to an asymptotical value at higher temperature. 
However, the maximum of $M_{\text{IMF}}$ is at $T\sim12$ MeV. 
Since no boundary condition is used in the present calculation, the density of the system will change with time
quickly,  especially for the source at high density.
For low density $\rho=0.05$ fm$^{-3}$, the initial distance between nucleons is large and the system will
 be easy to form a gas state at $T>0$ with no boundary condition, so the maximum of $M_{\text{IMF}}$ is close to $T=0$.
Thus it will be very interesting to study $\eta$ and $\eta/s$ for infinite uniform nuclear matter. 
In this case, $\eta/s$ and $M_{\text{IMF}}$ will be obtained at the same density condition. 

In summary, we have studied the shear viscosity of hot nuclear sources
with $0<T<30$ MeV at different densities. $\eta$ is 
calculated by the mean free path method in which the nucleon-nucleon 
collision number per fm/c plays an important role.
The absolute value of $\eta$ and its temperature dependence
is consistent with the previous studies. At $\rho\sim\frac{1}{2}\rho_0$, 
a minimum of $\eta/s$ is seen at around 10 MeV and a maximum of $M_{\text{IMF}}$ is also observed 
at the same temperature which is the indication of the liquid-gas phase transition.  
Thus a minimum of $\eta/s$ is observed at the location of the liquid-gas phase 
transition for $\rho\sim\frac{1}{2}\rho_0$, which is in accordance with the general phenomenon 
observed for other matter. To fully understand the relation between  $\eta/s$ and liquid-gas phase 
transition, further investigations on $\eta$ and $\eta/s$ for infinite uniform nuclear matter are expected. 

This work is supported by the Major State Basic Research Development Program of China 
under contract No. 2013CB834405, National Natural Science Foundation of China under 
contract No.s 11175231 and 11035009, and Knowledge Innovation Project of CAS under 
Grant No. KJCX2-EW-N01.

\end{document}